\documentclass[fleqn,10pt]{wlscirep}
\usepackage{url}
\usepackage[utf8]{inputenc}
\usepackage[T1]{fontenc}
\usepackage{amsmath}
\hypersetup{
    colorlinks = true,                    
    citecolor = {blue},
    linkcolor = {purple},
           }

\title {Containment efficiency and control strategies for the Corona pandemic costs}
\date{\today}

\author[1,*]{Claudius Gros}
\author[1]{Roser Valenti}
\author[1]{Lukas Schneider}
\author[2]{Kilian Valenti}
\author[3,4]{Daniel Gros}

\affil[1]{Institute of Theoretical Physics, Goethe University,
  60438 Frankfurt a.M., Germany}
\affil[2]{Vivantes Klinikum Spandau, 13585 Berlin, Germany}
\affil[3]{Department of Economics, University of California, Berkeley, USA}
\affil[4]{CEPS (Centre for European Policy Studies), 1000 Brussels, Belgium}

\affil[*]{Correspondence to gros@itp.uni-frankfurt.de}

\begin{abstract} 
  The rapid spread of the Coronavirus (COVID-19)
  confronts policy makers with the problem of measuring
  the effectiveness of containment strategies, balancing
  public health considerations with the economic
  costs of social distancing measures. We introduce a modified epidemic
  model that we name the controlled-SIR model, in which the disease reproduction
  rate evolves dynamically in response to political and societal
  reactions. An analytic solution is presented. The model
  reproduces official COVID-19 cases counts of a large number of
  regions and countries that surpassed the first peak of the outbreak.
  A single unbiased feedback parameter is extracted from field
  data and used to formulate an index that measures the efficiency
  of containment strategies (the CEI index). CEI values for a
  range of countries are given. For two variants of the controlled-SIR model,
  detailed estimates of the total medical and socio-economic
  costs are evaluated over the entire course of the epidemic. Costs
  comprise medical care cost, the economic cost of social distancing,
  as well as the economic value of lives saved. Under plausible
  parameters, strict measures fare better than a hands-off policy.
  Strategies based on current case numbers lead to substantially higher
  total costs than strategies based on the overall history of
  the epidemic.
  \end{abstract}

\begin{document} 

\flushbottom
\maketitle

\pdfbookmark[1]{Introduction}{Introduction}
\section*{Introduction}

In March 2020 the World Health Organization (WHO) declared
the Coronavirus (COVID-19) outbreak a pandemic~\cite{WHO2020}.
In response to the growth of infections and in particular
to the exponential increase in deaths~\cite{baud2020real},
a large number of countries were put under lockdown,
leading to an unprecedente recession~\cite{IMFoutlook2020}
which could potentially have longer term costs~\cite{adam2020specialEconomy}. In this
situation it is paramount to provide scientists, the
general public and policy makers with reliable estimates
of both the efficiency of containment measures ({\it{e.g.}} social distancing
and non-pharmaceutical health interventions), and
the overall costs resulting from alternative strategies.

\begin{figure*}[t!]
\centerline{
\includegraphics[width=0.9\textwidth]{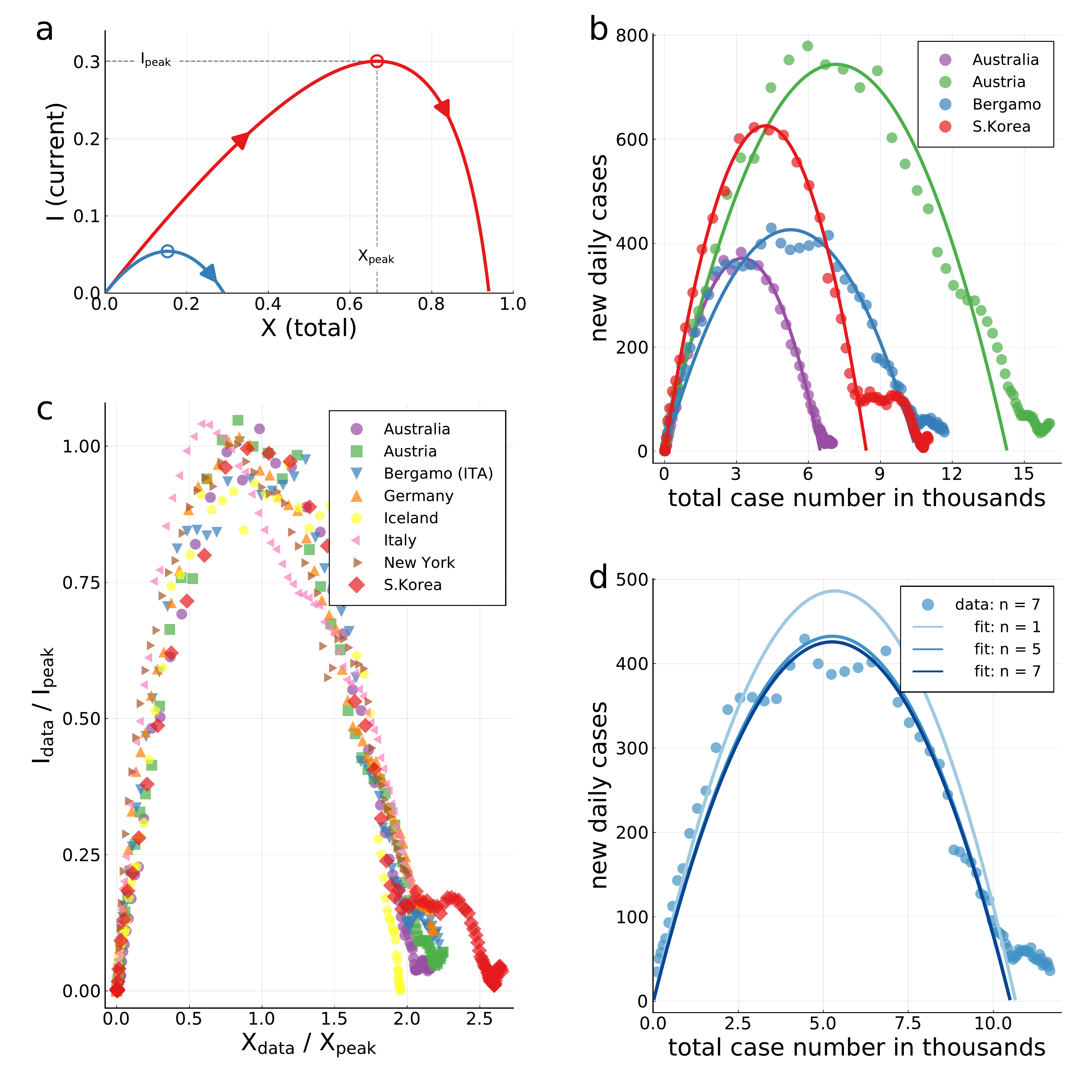}
           }
\caption{{\bf XI representation of COVID-19 outbreaks.}
{\bf a},  Model illustration. The closed phase-space expression 
$I=I(X)$ of actual infected cases $I$ as a function of total infected
cases $X$, as given by equation (\ref{XI}), is shown for two
cases: $\alpha_X=0$ (no control, red line) and $\alpha_X=10$
(long-term control, blue line) for an intrinsic reproduction
factor  of $g_0=3$. The
number of infections is maximal at $I_{\rm peak}$ (open circle),
after starting at $X=I=0$, with the epidemic ending when the
number of actual cases drops again to zero. At this point
the number of infected reaches $X_{\rm tot}$. The peak
$X_{\rm peak}=2/3$ of the uncontrolled case, $\alpha_X=0$, 
is sometimes called the `herd immunity' point. The final 
fraction of infected is $X_{\rm tot}=0.94$.
{\bf b}, Model validation for a choice of four countries/regions.
The model (lines) fits the
seven-day centered averages of COVID-19 case counts well. For
South Korea data till March 10 (2020) has been used for the
XI-fit, at which point a transition from overall control 
to the tracking of individuals is observable.
{\bf c}, Data collapse for ten countries/regions. 
Rescaling with the peak values
$X_{\rm peak}$ and $I_{\rm peak}$, obtained from the XI fit,
maps COVID-19 case counts approximately onto a universal
inverted parabola.
{\bf d}, Robustness test. The often strong daily fluctuations
are smoothed by $n$-day centered averages. Shown are the
Bergamo data (dots, $n=7$) and XI-fits to $n=1$ (no average),
$n=5$ and $n=7$. Convergence of the XI-representation is observed.
}
\label{fig_SIR_all}
\end{figure*}

The societal and political response to a major outbreak
like COVID-19 is highly dynamic, changing often rapidly
with increasing case numbers. We propose to model the
feedback of spontaneous societal and political reactions
by a standard epidemic model that is modified in one key
point: the reproduction rate of the virus is not
constant, but evolves over time alongside with the
disease in a way that leads to a `flattening of the curve'
\cite{branas2020flattening}. The
basis of our investigation is the SIR
(Susceptible, Infected, Recovered) model, which
describes the evolution of a contagious disease for
which immunity is substantially longer than the time-scale
of the outbreak \cite{kermack1927}. A negative
feedback-loop between the severity of the outbreak
and the reproduction factor $g$ is then introduced.
As a function of the control strength $\alpha_X$,
which unites the effect of individual, social and
political reactions to disease spreading, the
difference between an uncontrolled epidemic ($\alpha_X=0$) 
and a strongly contained outbreak (large $\alpha_X$) is 
described, as illustrated in
Fig.\,\ref{fig_SIR_all}a. The model, which we name
controlled-SIR model due to the presence of the
control parameter $\alpha_X$, is validated
using publicly available COVID-19 case counts from
a large range of countries and regions. We provide
evidence for data collapse when case counts of distinct
outbreaks are rescaled with regard to their peak values.
A comprehensive theoretical description based on an
analytic solution of the controlled-SIR model is given.
One finds substantial differences in the country-specific
intrinsic reproduction factor and its doubling time. The
controlled-SIR model allows in addition to formulate
an unbiased benchmark for the effectiveness of containment
measures, the containment efficiency index (CEI).

The controlled-SIR model is thoroughly embedded in
epidemiology modeling. Early on, the study
of the dynamics of measles epidemics \cite{bjornstad2002dynamics}
has shown that human behavior needs to be
taken into account \cite{funk2010modelling,bauch2013social}.
In this regard, a range of extensions to the
underlying SIR model have been proposed,
such as including the effect of vaccination,
contact-frequency reduction and quarantine
\cite{del2005effects}, human mobility
\cite{meloni2011modeling}, self-isolation
\cite{epstein2008coupled}, the effects
of social and geographic networks \cite{pastor2015epidemic}, 
the effects of awareness diffusion and epidemic propagation 
\cite{xia2012sir,wang2019impact}, and the influence
of explicit feedback loops~\cite{fenichel2011adaptive}.
For an in-depth description, epidemiology models need
to cover a range of aspects \cite{adam2020specialSimulations},
as the distinction between symptomatic and asymptomatic
cases \cite{chang2020modelling}, which prevents in general
the possibility of an explicit analytic handling. In the present 
work we pursue the alternative approach of retaining a minimal 
set of parameters, such that the resulting epidemiological 
model allows for an analytical description of the pandemic 
and its socio-economical aspects.

Political efforts to contain the pandemic, as social-distancing
measures and non-pharmaceutical health interventions,
are included in the controlled-SIR model as a dampening feedback 
mechanism. The controlled-SIR model is therefore suitable
to estimate the overall economic and health-related costs 
associated with distinct containment strategies, in particular 
when accumulated over the entire course of an epidemic outbreak. 
This approach, which is followed here,  extends classical 
studies of the economic aspects of controlling contagious 
diseases. A central question regards in this context the 
weighting of the economic costs of containment against the
cost of treatment, and the loss of life \cite{roberts2010guide,althouse2010public}.
For the value of life, statistical approaches attribute 
suitably estimated monetary values to an avoided premature death
\cite{murphy2006value,ashenfelter2004using,viscusi2003value}.
The resulting framework has been applied to the Corona pandemic
in several recent contributions in which the evolution of the
epidemic has been taken in general as exogenous
\cite{flatteningCurve2020}, relying on estimates for the
infection~\cite{ferguson2020impact} and case fatality 
rates \cite{rocklov2020covid,raoult2020coronavirus}.
Further studies have discussed the relative effectiveness of
control measures \cite{ferguson2020impact,wilder2020can,gatto2020spread,ferretti2020quantifying,chinazzi2020effect},
and the possible future course of the disease \cite{wilson2020modelling,tang2020estimation}.

\pdfbookmark[1]{Results}{Results}
\section*{Results}

\pdfbookmark[2]{Controlled-SIR Model}{controlled-SIR_model}
\subsection*{Controlled-SIR Model}
\begin{figure*}[t!]
\centerline{
\includegraphics[width=0.9\textwidth]{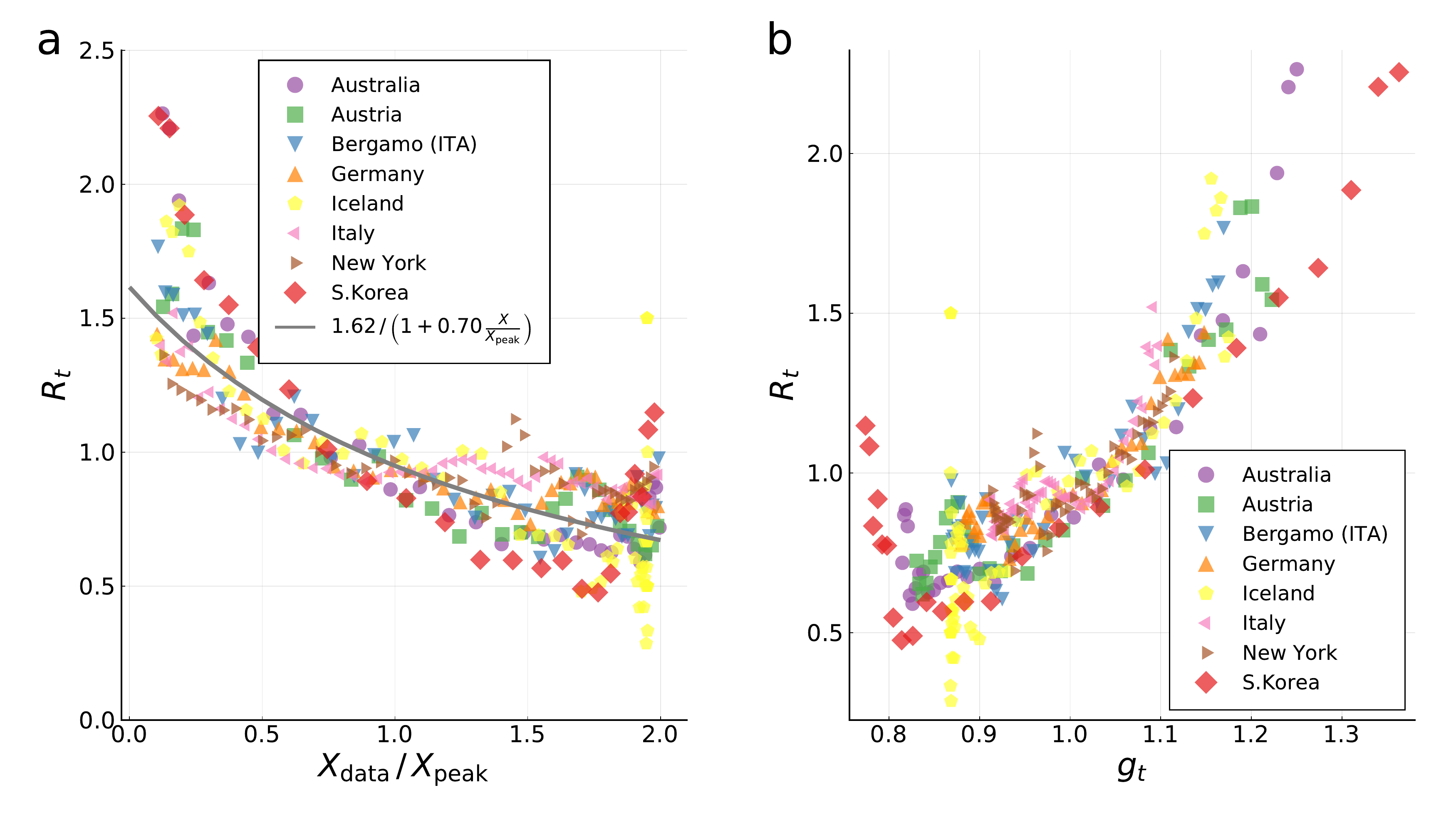}
           }
\caption{{\bf Validation of controlling feedback loop.}
The fraction of newly infected at time $t$ and at $t-4$ 
is used to estimate the time dependent reproduction factor 
$R_t = \overline{I}_t / \overline{I}_{t-4}$, when assuming 
a serial interval of four days (compare \cite{cori2013new}). 
Note that a seven-day centred moving average 
$\overline{I}_t = \sum_{s=t - 3}^{t+3} I_s$ is utilised.
\textbf{a,} $R_t$ as a function of the relative cumulative 
number $X / X_{\rm peak}$ of cases. A fit to the same functional 
form as in equation~(\ref{gSX}) is given (grey line). \textbf{b,} 
correlation of $R_t$ and $g_t$. The estimated reproduction 
factor $R_t$ is compared to the effective reproduction 
factor $g_t$ as defined in equation~(\ref{gSX}).
In \textbf{a} and \textbf{b} only data between 
$0.1 \le X / X_{\rm peak} \le 2$ is shown, with the lower
bound discarding the strong fluctuations in the early stages 
of the pandemic. The upper bound is used to define the termination
of the first wave.
}
\label{fig_validation}
\end{figure*}

In the following we introduce the model.
At a given time $t$ we denote with $S=S(t)$ the fraction
of susceptible (non-infected) individuals and $I=I(t)$
the fraction of the population that is currently ill
(active cases). Infected individuals can either recover 
or die as a consequence of the infection, here
we subsume both outcomes under $R=R(t)$, which denotes
hence the fraction of recovered or deceased individuals.
Normalization demands $S+I+R=1$ at all times. The
continuous-time SIR model \cite{gros2015complex}
\begin{equation}
\tau\dot S = -gSI,
\quad\quad
\tau\dot I = (gS-1)I,
\quad\quad
\tau\dot R = I
\label{SIR}
\end{equation}
describes an isolated epidemic outbreak characterized
by a timescale $\tau$ and a dimensionless reproduction
factor $g$. Social and political reactions reduce the
reproduction factor below its intrinsic (medical
disease-growth) value, $g_0$. We describe this functionality
as
\begin{equation}
g = \frac{g_0}{1+\alpha_X X}, \qquad\quad X=1-S\,,
\label{gSX}
\end{equation}
where we generalized standard epidemiological approaches
to nonlinear incidence rates
\cite{capasso1978generalization,hethcote1991some}.
The reaction to the epidemic is taken to be triggered by the
total fractional case count $X$ (i.e.\ the sum of active,
recovered and deceased cases), with $\alpha_X$ encoding the
reaction strength. In the Methods section we show how this 
functionality is validated by COVID-19 data, see also
Fig.\,\ref{fig_validation}.
In this view
$\alpha_X$ sums up the effects of an extended number of social
processes and political action taking. Further below we will
examine in addition strategies for which the response is
based on the fraction of actual active cases, $I$. We note
that containment due to a reduction in the reservoir of
susceptible $S$, is of minor importance, given that
COVID-19 infection cases are generally small with
respect to the overall population size.

The inverse functionality in equation~(\ref{gSX}) captures
the law of diminishing returns, namely that it becomes
progressively harder to reduce $g$ when increasing social
distancing. In this view, small reductions of $g$ are comparatively easy,
however a suppression by several orders of magnitude requires
 a near to total lockdown.
We denote equation (\ref{SIR}) together
with (\ref{gSX}) the controlled-SIR model.
Key to our investigation is the observation
that one can integrate the controlled-SIR
model analytically, as shown in the Methods
section, to obtain the phase-space relation
\begin{equation}
I = \frac{\alpha_X+g_0}{g_0}\,X+
\frac{1+\alpha_X}{g_0}\,\log(1-X)\,.
\label{XI}
\end{equation}
This relation, which we denote the
`XI representation', is manifestly independent
of the time scale $\tau$.

The medical peak load $I_{\rm peak}$ of actual infected
cases is reached at a total fractional case count
$X= X_{\rm peak}$, which is given by
\begin{equation}
gS=1, \qquad\quad
X_{\rm peak} =\frac{g_0-1}{g_0+\alpha_X}\,,
\label{Xpeak}
\end{equation}
For the case that $\alpha_X=0$ (no control),
$X_{\rm peak}$ reduces to the well-known result
$X_{\rm peak} =(g_0-1)/g_0$.

For finite $\alpha_X$, $I_{\rm peak}$ is obtained from
equations (\ref{XI}) and (\ref{Xpeak}),
\begin{equation}
I_{\rm peak} = \frac{g_0-1}{g_0} +
\frac{1+\alpha_X}{g_0}\,\log\left(
\frac{1+\alpha_X}{g_0+\alpha_X}\right)\,.
\label{I_peak}
\end{equation}
For $\alpha_X=0$, $I_{\rm peak}$ is sometimes called the
'herd immunity point'. The XI representation can be
parameterized consequently either by $g_0$ and $\alpha_X$,
as in equation (\ref{XI}), or indirectly by
$X_{\rm peak}$ and $I_{\rm peak}$, which are measurable
(modulo undercounting). In Fig.\,\ref{fig_SIR_all}a
an illustration of the XI-representation is given.
For $g_0=3$  and $\alpha_X=0$
one has $X_{\rm peak}=2/3$ and $I_{\rm peak}\approx 0.3$.
The total fraction of infected $X_{\rm tot}$ is
94\%, which implies that only about 6\% of the population
remains unaffected. Containment policies,  $\alpha_X>0$,
reduce these values.  Fig.\,\ref{fig_SIR_all}a and
equation (\ref{I_peak}) illustrate a sometimes encountered
misconception regarding the meaning of the herd immunity
point, which we have labeled simply $I_{\rm peak}$.
The epidemic doesn't stop at $I_{\rm peak}$ since infections
continue beyond this point, albeit at a declining rate.

\pdfbookmark[2]{XI representation of COVID-19 outbreaks}{XI_representation}
\subsection*{XI representation of COVID-19 outbreaks}
In Fig.\,\ref{fig_SIR_all}b,c we show for a representative 
choice of countries, regions and cities that COVID-19 
outbreaks are described by the controlled-SIR model 
to an remarkable degree of accuracy. For the analysis 
presented in Fig.\,\ref{fig_SIR_all}b,c
we divided, as described in the Methods section,
the official case counts by the nominal population
size of the respective region or country. Seven-day
centered averages are performed in addition. The 
country- and region-specific XI representations are 
then fitted by equation (\ref{XI}). The fact that 
the outbreaks are well described by the model, 
independently of the size of the country, region 
or city, evidences the applicability of the
controlled-SIR model.

It has been widely discussed that official case
counts are affected by a range of factors,
which include the availability of testing facilities
and the difficulty to estimate the relative
fraction of unreported cases \cite{lachmann2020correcting,li2020substantial}.
For example, as of mid-March 2020, the degree of testing for
COVID-19, as measured by the proportion of the entire population,
varied by a factor of 20 between the United States
(340 tests per million) and South Korea (6100 tests per
million)~\cite{owidcoronavirus}. The true incidence might
be, according to some estimates~\cite{qiu2020covert}
higher by up-to a factor of ten than the numbers
reported in the official statistics as positive.

Case counts enter the XI representation in both 
the $x-$ and $y-$ axis. Scaling both $I$ and $X$ 
with a constant factor allows therefore to 
compensate for the undercounting problem. At
the same time the control strength $\alpha_X$ needs
to be rescaled, a procedure implicitly implemented
for the fits shown in Fig.\,\ref{fig_SIR_all}b,c. The XI
framework is in this sense robust. Renormalization
becomes however invalid if the undercounting of
infection cases changes abruptly at a certain point during
the epidemics, f.i.\ as a result of substantially
increased testing. We will come back to this point
further below. A fundamental change in the strategy
followed by the government, e.g.\ from laissez faire
to restrictive, would lead likewise to a change
in $\alpha_X$, which is not captured in the current
framework.

In the analysis presented in Fig.\,\ref{fig_SIR_all}
daily case counts were taken as proxies for the number (relative fraction),
 of infected individuals
$I=I(t)$. This assumption holds only up to a rescaling
factor, which implies that the $g_0$ extracted for a given
country or region is not the native, but an effective
reproduction factor. To see this consider, e.g., the
initial slope, $I\sim X(g_0-1)/g_0 $, as given by
equation~(\ref{slope_X}). Rescaling daily case counts
in order to obtain estimates for the number of infected individuals
changes the slope and hence $g_0$. Given that the appropriate
rescaling of daily case counts can only be estimated, and that
we are interested here in a simple but accurate effective modeling
of COVID-19 outbreaks, and not in the extraction of the native
reproduction factor, we did not pursue this route.

In Table~\ref{table_CEI} we present for a number of
countries and regions the obtained effective growth
factors $g_0$ and the corresponding doubling
times $\tau_2$, where $\tau_2=\log(2)/\log(g_0)$
defines the number of time units $\tau$ needed to
double case numbers. As expected, according to the 
description above, one finds that the values of $g_0$ 
are substantially lower than the consensus
estimates 2-3 for the native reproduction number
\cite{leung2020first,kucharski2020early,wu2020nowcasting,alimohamadi2020,yuan2020monitor}. The observed
doubling times $\tau_2$ are however retained when adapting
the effective time scale $\tau$ accordingly.

For a robustness check we evaluated the parameters of the
controlled-SIR model assuming that only a fraction $f$ of
the nominal population of the country or region in question
could be potentially infected, possibly
due to the presence of social or geographical barriers
to the disease spreading. Only marginal differences
were found for $f=1/3$. The data presented in
Table~\ref{table_CEI} suggest most countries followed
in the first wave of the COVID-19 pandemic
strict containment policies, as measured in terms of
the CEI index. This insight is of particular relevance
for the  discussion of the costs incurring for the various
containment strategies presented further below.

\pdfbookmark[2]{Data collapse for COVID-19}{Data_collapse}
\subsection*{Data collapse for COVID-19}
Given that the XI representation is determined solely
by two quantities, $X_{\rm peak}$ and $I_{\rm peak}$,
universal data collapse can be attained by plotting
field data normalized with regard to the respective
peak values, viz by plotting $I/I_{\rm peak}$ as a
function of $X/X_{\rm peak}$. It is remarkable, to
which degree the country- and region specific official
case counts coincide in relative units, see
Fig.\,\ref{fig_SIR_all}c. It implies that the
controlled-SIR model constitutes a faithful
phase-space representation of epidemic spreading
subject to socio-political containment efforts.

\begin{table}[b!]
\caption{{\bf COVID-19 containment efficiency index.}
For selected countries/ regions, key COVID-19 parameters, as
extracted from the respective official case counts.
Given is the dimensionless reproduction factor $g_0$,
the doubling time $\tau_2=\log(2)/\log(g_0)$, in units of
$\tau$, and the containment efficiency index
$\mbox{CEI}=\alpha_X/(g_0+\alpha_X)$. Note that $g_0$
is not the native, but an effective reproduction factor.
\label{table_CEI}
}
\centerline{\setlength\arrayrulewidth{1pt}
\begin{tabular}{lcc|crcrcr}
location &&&& $g_0$ && $\tau_2$ && CEI \\ \hline
Italy & ITA &&& 1.17 &&  4.4 && 0.991\\
Iceland & ISL &&& 1.19 &&  4.0 && 0.983\\
Bergamo & ITA &&& 1.20 &&  3.8 && 0.972\\
Roma & ITA &&& 1.20 &&  3.8 && 0.998\\
Germany & DEU &&& 1.21 &&  3.6 && 0.995\\
United States & USA &&& 1.22 &&  3.5 && 0.994\\
Spain & ESP &&& 1.23 &&  3.3 && 0.990\\
Luxembourg & LUX &&& 1.28 &&  2.8 && 0.988\\
Austria & AUT &&& 1.30 &&  2.6 && 0.997\\
Israel & ISR &&& 1.30 &&  2.6 && 0.997\\
Australia & AUS &&& 1.32 &&  2.5 && 0.999\\
South Korea & KOR &&& 1.46 &&  1.8 && 1.000\\
\end{tabular}}
\end{table}

\pdfbookmark[2]{Asymmetry of up-/down time scales}{asymmetry}
\subsection*{Asymmetry of up-/down time scales}
For the controlled SIR model an explicit analytic expression
for the $X-I$ phase space representation can be derived,
as given by equation (\ref{XI}), but not for the complete
timeline $X(t)$ and $I(t)$. Exploiting the fact that
case counts are generally small with respect to the population
for real-world epidemic outbreaks, the universal relation
\begin{equation}
\frac{\mbox{time down from the peak}}{\mbox{time up to the peak}}
= 2g_0-1
\label{time_down_up}
\end{equation}
between the time the outbreak needs to retreat from the peak,
and to reach it in first place, can however be found, as shown
in the Methods section. Interestingly, the ratio of down-/ and
up-times is independent of the control strength $\alpha_X$
(if and only if $X\ll1$), which suggests that equation (\ref{time_down_up})
is valid for epidemic outbreaks in general.
For COVID-19, typical values of the effective $g_0$ are of the order of 1.2-1.3,
as listed in Table~\ref{table_CEI}, which implies that
outbreaks take of the order of 40-60\% longer to retreat
than to ramp up.

\pdfbookmark[2]{Containment efficiency index}{CEI}
\subsection*{Containment efficiency index}
The control strength $\alpha_X$ enters the reproduction
factor as $\alpha_X X$, see equation~(\ref{gSX}). Data collapse
suggest that regional and country-wise data is comparable
on a relative basis. From
$\alpha_X X=(\alpha_X X_{\rm peak})(X/X_{\rm peak})$
it follows that $\alpha_X X_{\rm peak}=\alpha_X(g_0-1)/(g_0+\alpha_X)$
is a quantity that measures the combined efficiency
of socio-political efforts to contain an outbreak.
Dividing by $g_0-1$ results in a normalized index,
the `Containment Efficiency Index' (CEI):
\begin{equation}
\mbox{CEI} = \frac{\alpha_X X_{\rm peak}}{g_0-1} = \frac{\alpha_X}{g_0+\alpha_X}\,,
\label{CEI}
\end{equation}
with $\mbox{CEI}\in[0,1]$. The index is unbiased, being
based solely on case count statistics, and not on
additional socio-political quantifiers. Our estimates
are given in Table~\ref{table_CEI}. The values for
the evaluated regions/ countries are consistently
high, close to unity, the upper bound, indicating
that the near-to-total lockdown policies implemented by
most countries have been effective in containing
the spread of COVID-19. A somewhat reduced CEI value
is found for the particularly strongly affected Italian
region of Bergamo. For South Korea the CEI is so high that
its deviation from unity cannot be measured with confidence.

\begin{figure}[t!]
\centerline{
\includegraphics[width=0.6\columnwidth]{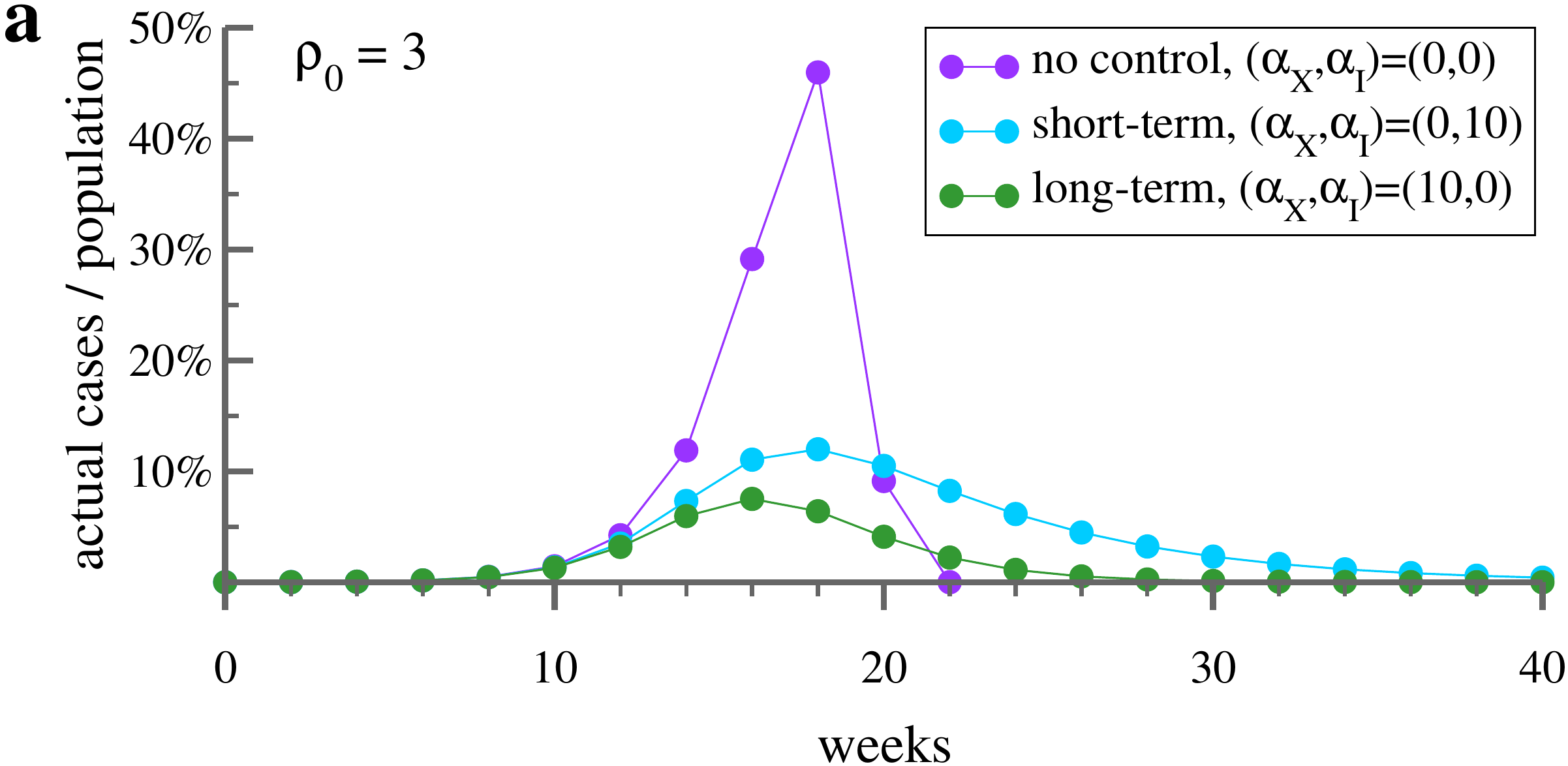}
           }
\centerline{
\includegraphics[width=0.6\columnwidth]{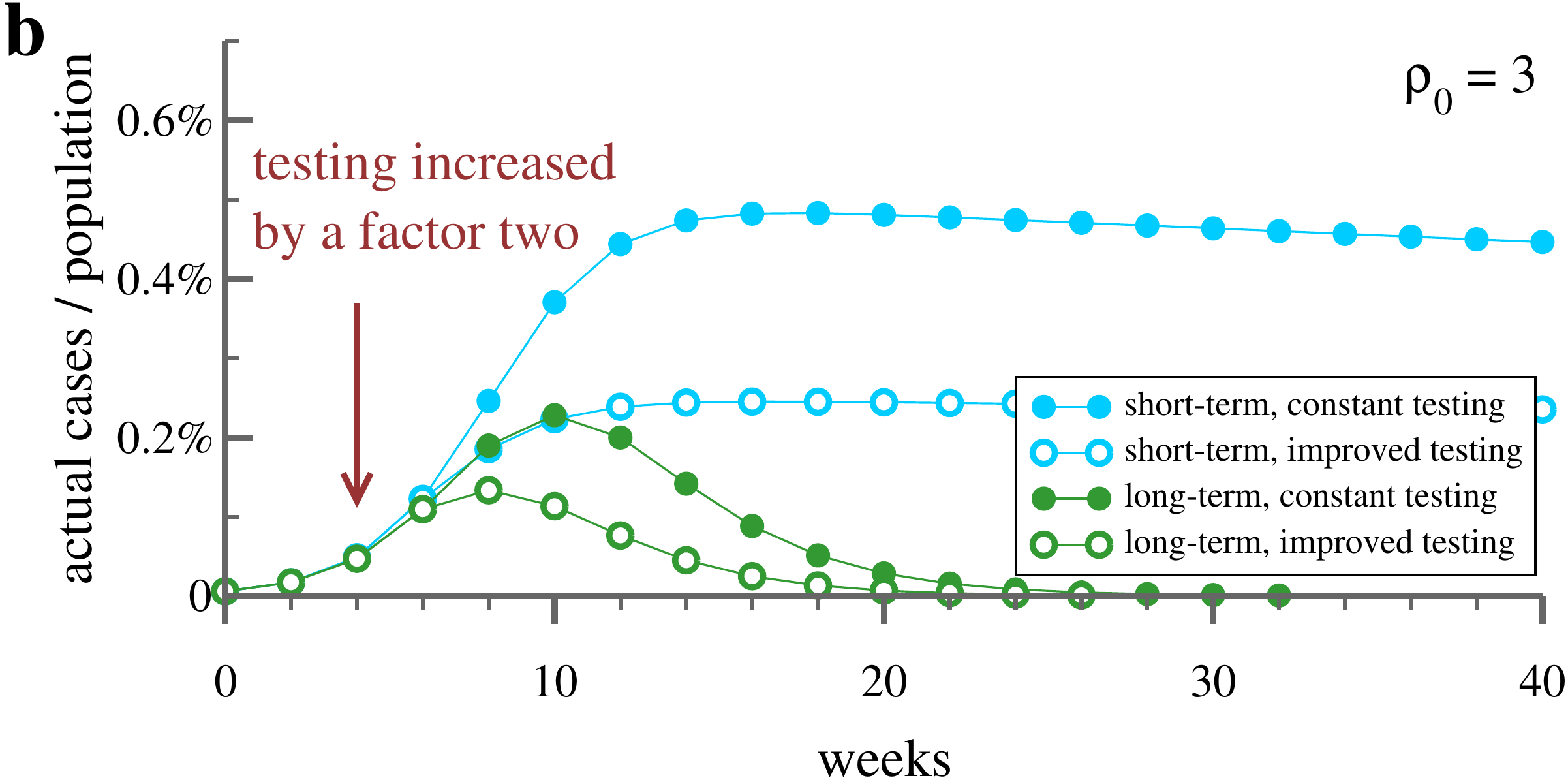}
           }
\caption{{\bf Control of epidemic peak.}
{\bf a},
Shown is the timeline of actual infected cases during an epidemic
outbreak with an intrinsic reproduction factor of $\rho_0=3.0$
	defined in the discrete model, which
is close to COVID-19 estimates \cite{liu2020reproductive}.
The simulation is obtained by iterating equation~(\ref{SIR_logistic}),
with one iteration corresponding
to two weeks, taken as the average duration of the illness.
Short-term control, when responding to the actual
number of cases, see equation~(\ref{gIX}), is able to
reduce the peak strain on the hospital system, but only
by prolonging substantially the overall duration. Long-term
control, which takes the entire history of the
outbreak into account, is able to reduce both the peak
and the duration of the epidemic.
{\bf b},
Increasing testing by a factor two (arrow), reduces the
undercounting factor which increases, in turn, the effective
response strength for both, the peak number of actual
cases and the duration of the outbreak. Here
$(\alpha_X,\alpha_I)=(400,0)\,/\,(0,400)$ has been
used respectively for long-\,/\,short-term control.
}
\label{fig_time_evolution}
\end{figure}

\pdfbookmark[2]{Long-term vs.\ short-term control}{short-term_control}
\subsection*{Long-term vs.\ short-term control}
So far, in equation (\ref{gSX}) it was assumed that society
and policy makers react to the total case count of infected $X$.
This reaction pattern, which one may denote as `long-term control',
describes field data well. It is nevertheless of interest
to examine an alternative, short-term control:
\begin{equation}
g = \left\{\begin{array}{lcl}
g_0/(1+\alpha_I I) && \mbox{(short-term)}\\[0.5ex]
g_0/(1+\alpha_X X) && \mbox{(long-term)}\\
\end{array}\right.
\label{gIX}
\end{equation}
For short-term control the relevant yardstick is
given by the actual case number of infected $I$.
In reality, people will react to officially reported
case counts, which are affected by the undercounting
problem. For the terms $\alpha_I I$ and and $\alpha_X X$
in equation (\ref{gIX}) this corresponds to a
renormalization of reaction parameters $\alpha_I$ and $\alpha_X$.

\begin{figure}[t!]
\centerline{
\includegraphics[width=0.6\columnwidth]{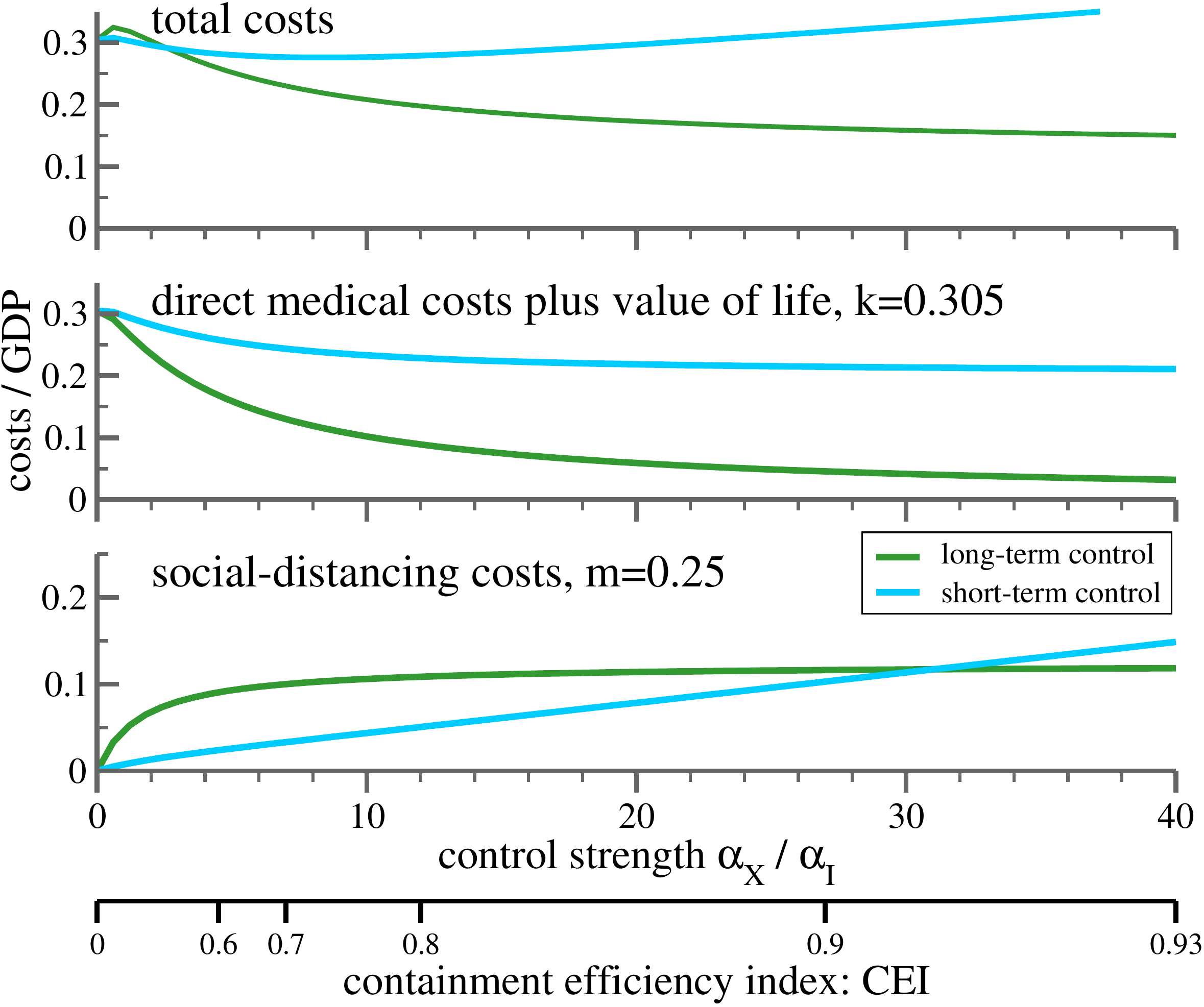}
           }
\caption{{\bf Cost of epidemic control strategies including value 
of life.} Shown are the costs in terms of GDP$_{\rm p.c.}$,
for long-term and short-term control,
as defined by equation~(\ref{gIX}), both as
a function of $\alpha_X$ and the CEI values (\ref{CEI}),
as indicated by the additional axis at the bottom.
Given are the costs incurring from social distancing,
equation~(\ref{costs_social}) with $m=0.25$ (lower panel),
the pure medical costs with value of life costs (middle  panel),
and the sum of social and medical costs (upper panel). It
is assumed that the containment policy switches from mass
control to individual tracking when the fraction of actual cases
$I_t$ drops below a threshold of $I_{\rm min}=10^{-5}$.
The starting $I_0=2\cdot10^{-5}$.
}
\label{SIR_costs_250_305}
\end{figure}

Both control types, short- and long-term, can be employed
either for the continuous-time SIR model, equation (\ref{SIR}),
or for the discrete-time variant,
\begin{equation}
I_{t+1} = \rho_t I_t (1-X_t), \quad\quad
X_t = \sum_{k=0}^{\infty} I_{t-k}\,,
\label{SIR_logistic}
\end{equation}
The time-dependent reproduction factor has
been denoted here as $\rho_t$, in order to
make clear that discrete times are used.
Short- and long-term control is then equivalent
to $\rho_t=\rho_0/(1+\alpha_I I)$ and
$\rho_t=\rho_0/(1+\alpha_X X)$. One time
step corresponds for the discrete-time SIR
model to the mean infectious period.

The simulations of equation (\ref{SIR_logistic})
presented in Fig.\,\ref{fig_time_evolution} illustrate
the capability of short-term and long-term reaction
policies to contain an epidemic. While both strategies
are able to lower the peak of the outbreak
with respect to the uncontrolled ($\alpha_X=\alpha_I=0$)
case, the disease will become close to endemic when
the reaction is based on the actual number of cases,
$I_t$, and not on the overall history of the outbreak.

Also included in the lower panel of Fig.\,\ref{fig_time_evolution}
is a protocol simulating an increase of testing by a factor
of two. Here $(\alpha_X,\alpha_I)=(400,0)$ and
$(\alpha_X,\alpha_I)=(0,400)$ have been used as the 
starting reaction strengths, respectively for long- 
and short-term control, which are increased by a 
factor of two when testing reduces the
undercounting ratio by one half. One observes that long-term
control is robust, in the sense that increased
testing contributes proportionally to the containment
of the outbreak. Strategies reacting to daily case
number are in contrast likely to produce an endemic
state.

The framework developed here, equations (\ref{SIR})
and (\ref{gSX}), describes mass control strategies,
which are necessary when overly large case numbers
do not allow to track individual infections. The
framework is not applicable once infection rates are
reduced to controllable levels by social distancing
measures. The horizontal 'tail' evident in the data
from South Korea in Fig.\,\ref{fig_SIR_all}b can be
taken as evidence of such a shift from long-term
mass control to the tracking of individual cases.

\pdfbookmark[2]{Costs of controlling the COVID-19 pandemic}{cost_controlled_pandamic}
\subsection*{Costs of controlling the COVID-19 pandemic}
As shown above, the controlled-SIR model  allows
for a faithful modeling of the entire course of
an isolated outbreak. We apply it now to investigate how
distinct policies and societal reaction patterns,
as embedded in the parameter $\alpha_X$, influence
the overall costs of the epidemic. This is an inter-temporal
approach since the cost of restrictions today to public
life (lockdowns, closure of schools, etc.) must be set
against future gains in terms of lower infections (less
intensive hospital care, fewer deaths).
Four elements
 dominate the cost structure:  (i) The
working time lost due to an infection, (ii) the direct
medical costs of infections, (iii) the value of life costs, and
(iv) the cost related to `social distancing'. The first three
are medical or health-related. All costs can be scaled in terms
of GDP per capita (GDP$_{\rm p.c.}$). This makes our analysis
applicable not only to the US, but to most countries with
similar GDP$_{\rm p.c.}$, e.g.\ most OECD countries.

\pdfbookmark[2]{Overall cost estimates}{overall_cost}
\subsection*{Overall cost estimates}
The cost estimates, which are given in detail in the
Supplementary Information, can be performed disregarding
discounting. With market interest rates close to zero
and the comparatively short time period over which the epidemic
plays out, a social discount rate between 3\% and 5\%
would make little difference over the course of one year
\cite{moore2004just}.

Total health costs $C^{\rm medical}$ incurring over the
duration of the epidemic are proportional to the overall
fraction $X_{\rm tot}=X_{t\to\infty}$ of infected, with
a factor of proportionality $k$. We hence have
$C^{\rm medical} =kX_{tot}$. We estimate $k\approx0.305$
in terms of  GDP$_{\rm p.c.}$
when all three contributions (working-time lost,
direct medical cost, value of life) are taken into account,
and $k\approx0.14$ when value of life costs are omitted.

The economic costs induced by social-distancing measures,
$C^{\rm social}$, depend in a non-linear way on the evolution
of new cases (short-term control) or the percentage
of the population infected (long-term control). To
be specific, we posit that the reduction of economic activity
is percentage-wise directly proportional to the relative
reduction in the reproduction factor \cite{gros2020economics}, 
viz to $(\rho_0-\rho_t)/\rho_0$:
\begin{equation}
C^{\rm social} = \sum_{I_{t}>I_{\rm min}} \mbox{c}_t^{\rm s},
\quad\quad
\mbox{c}_{t}^{\rm s} = m\
\frac{\rho_0-\rho_t}{\rho_0}\,\frac{2}{52}\,,
\label{costs_social}
\end{equation}
where $2/52$ is the per year fraction of 2-week quarantine
period. The epidemic is considered to be under control when
the fraction of new infections $I_t$ falls below a minimal
value $I_{\rm min}$. As detailed out in the Supplementary
Information, a comprehensive analysis yields $m\approx 0.25$ in
terms of GDP$_{\rm p.c.}$. Note that the ansatz equation
(\ref{costs_social}) holds only when mass control
is operative, viz when large case numbers do not
allow the tracking of individual infections.

\begin{figure}[t!]
\centerline{
\includegraphics[width=0.6\columnwidth]{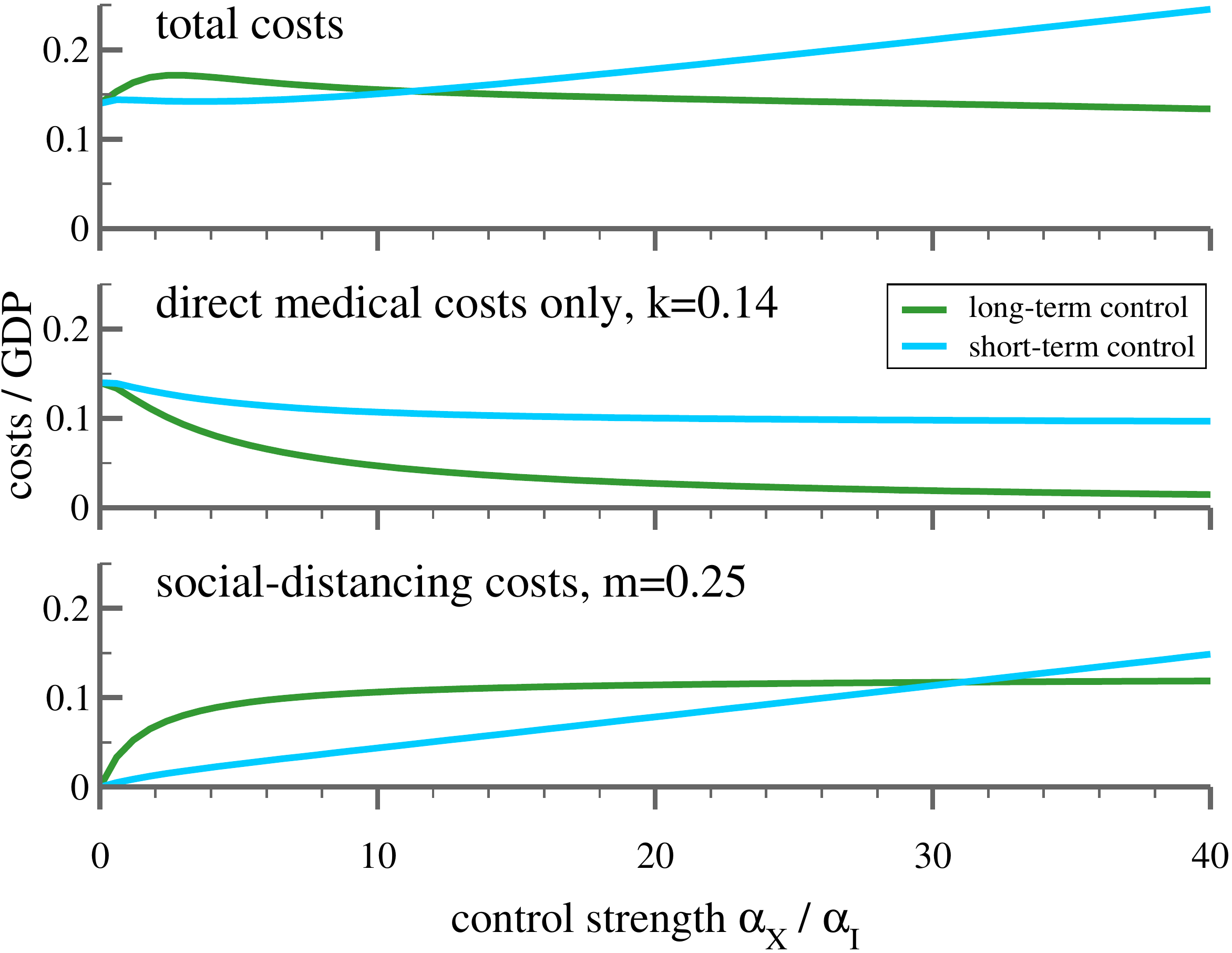}
           }
\caption{{\bf
Cost of epidemic control without value of life.}
As in Fig.\,\ref{SIR_costs_250_305} (bottom panels are identical),
but without the value of life costs. A long-term strategy with
intermediate reaction strength is costlier than a hands-off policy.
}
\label{SIR_costs_250_140}
\end{figure}

Once $k$ and $m$ are known, one can compare the total costs
incurring as the result of distinct policies by computing
the sum of future costs for different values for $\alpha_X$
in equation~(\ref{gSX}). This is illustrated in Fig.\,\ref{SIR_costs_250_305}
with the value of life costs included ($k=0.305$), and
in Fig.\,\ref{SIR_costs_250_140}, without
value of life costs ($k=0.14$). Given are the total
cumulative costs for the two strategies considered,
long-term and short-term control, both as a
function of the respective implementation strength,
as expressed by the value of $\alpha_X$ and $\alpha_I$.

The middle panel of Fig.\,\ref{SIR_costs_250_305}
shows that a society focused on short-term successes will
incur substantially higher medical costs, because restrictions
are relaxed soon after the peak. By contrast, if policy
(and individual behavior) is influenced by the total
number of all cases experienced so far, restrictions will not be
relaxed prematurely and the medical costs will be lower for
all values of $\alpha_X$. The bottom panel shows the
social distancing costs as a fraction of GDP$_{\rm p.c.}$, which
represent a more complicated trade-off between the
severity of the restrictions and the time they need
to be maintained. If neither policy, nor individuals
react to the spread of the disease ($\alpha_X=0$) the
epidemic will take its course and costs are solely medical.
This changes as soon as  society reacts, i.e.\ as
$\alpha_X$ increases. Social distancing costs increase
initially (i.e.\ for small values of $\alpha_X$), somewhat
stronger for the long-term than for the short-term
reaction framework. The situation reverses for higher values
of $\alpha_X$ and $\alpha_I$ with $\alpha_X, \alpha_I
\approx30$ being the turning point.
From there on, the distancing cost from a long-term based
reaction falls below that of the short-term strategy.
The sum of the two costs is shown in the uppermost panel.
For large values of $\alpha_X$, $\alpha_I$ short-term policies result
in systematically higher costs.

Supplementary Figure 1 of the Supplementary Information shows
that short-term control cannot explain observed COVID-19
outbreaks per se. Our estimates for the incurring costs
suggest that economic cost considerations may
have caused countries to follow predominantly
long-term control strategies during the first wave of the
COVID-19 outbreak.

\pdfbookmark[1]{Discussion}{Discussion}
\section*{Discussion}

The total costs of competing containment
strategies can be estimated if the feedback
of socio-political measures can be modeled.
For this one needs two ingredients: (i) a validated
epidemiological model and (ii) a link between
the impact of containment efforts, in terms of
model parameters, to their economic costs.
Regarding the first aspect, we studied the
controlled-SIR model and showed that COVID-19
outbreaks follow in many cases the phase-space
trajectory, the XI representation, predicted
by the analytic solution. The same holds for the
2015 MERS outbreak in South Korea, as shown in
Fig.\,\ref{fig_alpha_0}b.
We extracted for a number of countries and
regions estimates for the intrinsic
doubling times and found that they are not
correlated to the severity of the outbreak.
Regarding the second aspect, we proposed that the
economic costs of social distancing are proportional
to the achieved reduction in the infection rate 
\cite{gros2020economics}. Equation (\ref{costs_social}) 
establishes the required link between epidemiology, political
actions and economic consequences. Health-related
costs, which are related to official case counts,
are in contrast comparatively easier to estimate.
We have not considered formally the optimal
control problem, which would consist of minimising the
sum of total costs if the control strength could be
chosen freely for every period. Instead, we have been
interested here in comparing distinct containment strategies
under which society and governments react in a predictable
pattern to the spread of the disease.

A non-trivial outcome of our study is that strong suppression
strategies lead to lower total costs than taking no action,
when containment efforts are not relaxed with falling infection
rates. A short-term control approach of softening containment
with falling numbers of new cases is likely to lead to a prolonged
endemic period. With regard to the `exit strategy' discussion,
these findings imply that social distancing provisions need
to be replaced by measures with comparative containment power.
A prime candidate is in this regard to ramp up testing
capabilities to historically unprecedented levels, several
orders of magnitude above pre-Corona levels. The epidemic
can be contained when most new cases are tracked,
as implicitly expressed by the factor $\alpha_X$.
This strategy can be implemented once infection rates are
reduced to controllable levels by social distancing measures.
Containment would benefit if the social or physical separation
of the `endangered' part of the population from the
`not endangered' would be organized in addition on a
country-wide level, as suggested by community-epidemiology.
With this set of actions the vaccine-free period can be bridged.

As a last note, there is a sometimes voiced misconception regarding
the meaning of the herd immunity point, which occurs for an infection
factor of three when 66\% of the population is infected.
Beyond the herd immunity point, the infected-case counts remain
elevated for a considerable time. The outbreak stops
completely only once 94\% of the population has been
infected, as illustrated in Fig.\,\ref{fig_SIR_all}a.



\pdfbookmark[1]{Methods}{Methods}
\section*{Methods}

\pdfbookmark[2]{Validation of the model from COVID-19 data}{validation}
\subsection*{Validation of the model from COVID-19 data}
In Fig.\,\ref{fig_validation} we show how the model given in
equation~(\ref{gSX}) is validated by COVID-19 data. 
Fig.\,\ref{fig_validation}a displays the collected data of 
infected population during the first wave of the COVID-19 pandemic 
in a range of representative countries and regions. Plotted 
is the time-dependent reproduction factor R$_t$ as a function 
of the relative cumulative case count $X / X_{\rm peak}$. We 
followed standard procedures \cite{cori2013new} and defined
$R_t$ as the fraction of newly infected individuals 
at time $t$ with respect to the infected individuals
at time $t-4$ days, $R_t = \overline{I}_t / \overline{I}_{t-4}$,
where seven-day centered moving averages 
$\overline{I}_t = \sum_{s=t - 3}^{t+3} I_s$ are considered.
Also shown is a fit to the data using the functional form 
predicted by our model, equation~(\ref{gSX}). The
quantitative comparison between field data and modeling
validates the controlled-SIR model. For a set of 
representative countries and regions it is shown
in Fig.\,\ref{fig_validation}b that there is a direct 
correlation between the measured reproduction factor 
$R_t$ and the effective reproduction factor $g_t$, 
as defined by equation~(\ref{gSX}). 

\pdfbookmark[2]{Data collection and handling}{data_collection}
\subsection*{Data collection and handling}
Data has been accessed as of May 18 (2020) via the public
COVID-19 Github repository of the Johns Hopkins Center
of Systems Science and Engineering \cite{github2020}.
Preprocessing was kept minimal, comprising only a basic
smoothing with sliding averages. If not stated otherwise,
a seven time centered average (three days before/after, 
plus current day) has been used. Robustness checks with 
one, three and five day sliding averages were performed, 
as shown in Fig.\,\ref{fig_SIR_all}d.
Fractional case counts are obtained by dividing the raw
number by the respective population size. For the case
of South Korea, the XI-analysis was performed using the initial
outbreak, up to March 10 (2020). China has been ommitted
in view of the change in case count methodolgy mid
February 2020.

The variable $I$ represents in the SIR model the fraction
of the population that is infectious, which for this model coincides
with the infected population. For the COVID-19 data,
we used instead an XI-representation for which the number of
new daily cases is plotted against the total case count. This
procedure is admissible as long as the relative duration of
the infectious period does not change.

\pdfbookmark[2]{Fitting procedure}{fitting}
\subsection*{Fitting procedure}
We compared the theoretical result for the controlled
SIR model, $I(X)\equiv I^{\rm(theory)}(X)$,
see equation (\ref{XI}), to the reported data
$I_t^{\rm(data)}$, where $t$ runs over all days.
The field data $X_t^{\rm(data)}$ for the total case number
is crowded at low levels of $X$ and $I$ in the XI
representation. A fitting procedure that takes the
range $X\in[0,X_{\rm tot}]$ uniformly into account
is attained when minimizing the weighted loss function
\begin{equation}
U = \sum_t u_t\left(I_t^{\rm(data)}-I^{\rm(theory)}(X_t^{\rm(data)})\right)^2\,.
\label{loss_function}
\end{equation}
For the  weight we used
$u_t=X_t^{\rm(data)}-X_{t-1}^{\rm(data)}= I_t^{\rm(data)}$,
which satisfies the sum-rule  $\sum_t u_t=X_{\rm tot}$,
where $X_{\rm tot}$ is the total (maximal) case count.
With equation (\ref{loss_function}) it becomes irrelevant
where the timeline of field data is truncated, both at the
start or at the end. Adding a large number of null
measurements after the epidemic stopped would not
alter the result. Numerically the minimum of $U$ as
a function of $g_0$ and $\alpha_X$ is evaluated.

\pdfbookmark[2]{Modeling field data as uncontrolled outbreaks}{modeling_field_data}
\subsection*{Modeling field data as uncontrolled outbreaks}
\begin{figure}[t!]
\centerline{
\includegraphics[width=0.9\columnwidth]{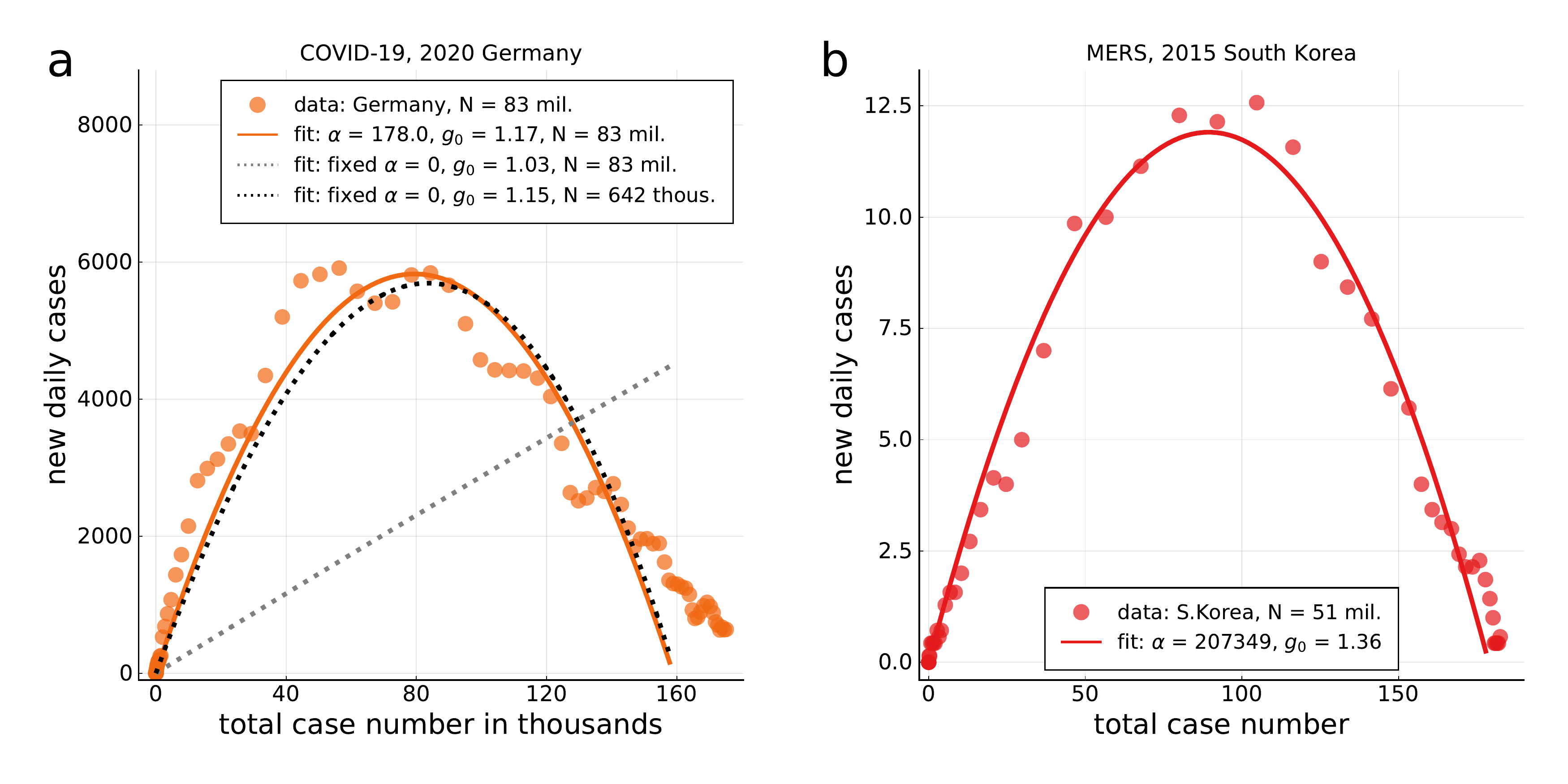}
           }
\caption{{\bf Case count modelling.}
{\bf a}, Modeling case counts as uncontrolled outbreaks.
Case counts, here for Germany (seven-day centered averages, dots),
can be modeled using either the full XI representation (full line),
as given by equation (\ref{XI}), or with the standard uncontrolled
SIR model ($\alpha_X=0$, dashed lines). Using the nominal population
size for Germany, 83 Million, leads to an utterly unrealistic
$\alpha_X=0$ curve (dashed, grey). The best $\alpha_X=0$ fit is obtained
when a fictitious population size of 478 Thousand is assumed
(dashed, black). An epidemic abates on its own only when
the population size is of the order of the total case count
divided by $X_{\rm tot}$.
{\bf b}, XI representation of the 2015 MERS outbreak in
South Korea, covering a total of 186 cases. A $n=7$
centered average has been used, in view of the small
case numbers.
}
\label{fig_alpha_0}
\end{figure}

It is of interest to examine to which degree official
case statistics could be modeled using an uncontrolled
model, $\alpha_X=0$. For this purpose it is necessary to
assume that the epidemics stops on its own, which implies
that one needs to normalize the official case counts
not with respect to the actual population, but with
respect to a fictitious population size $N$.
In this view the outbreak starts and ends in a socially
or geographically restricted community. The results
obtained when optimizing $N$ are included in
 Fig.\,\ref{fig_alpha_0}a. At first sight,
the $\alpha_X=0$ curve tracks the field data. Note however the
very small effective population sizes, which are found to
be 478000 for the case of Germany. Alternatively one
may adjust $g_0$ by hand during the course of an epidemic,
as it is often done when modeling field data.

\pdfbookmark[2]{Analytic solution of the controlled-SIR model}{analytic_solution}
\subsection*{Analytic solution of the controlled-SIR model}
Starting with the expression for the long-term control,
equation~(\ref{gSX}), one can integrate the controlled-SIR model
equation (\ref{SIR}) to obtain a functional relation between $S$ and $I$.
Integrating $\dot{I}/\dot{S}$, viz
\begin{equation*}
dI = -dS +\frac{1}{g(S)S}\,dS = -dS +
\frac{1}{g_0}\,\frac{1+\alpha_X(1-S)}{S}\,dS\,,
\end{equation*}
yields
\begin{equation}
I = -\left(\frac{\alpha_X}{g_0}+1\right)S+
\frac{1+\alpha_X}{g_0}\,\log(S) +c\,,
\label{I_S_c}
\end{equation}
where the integration constant $c$ is given by
the condition $I(S\!=\!1)=0$. Substituting
$S=1-X$ one obtains consequently the
XI-representation equation (\ref{XI}). The analogous result
for $\alpha_X=0$ has been derived earlier~\cite{harko2014exact}.
The number of actual cases,
$I$, vanishes both when $X=0$, the starting point of
the outbreak, and when the epidemic stops. The overall
number of cases, $X_{\rm tot}$, is obtained consequently
by the non-trivial root $X_{\rm tot}$ of equation (\ref{XI}), as
illustrated in Fig.\,\ref{fig_SIR_all}a.
As a side remark, we mention that the XI
representation allows us to reduce
equation (\ref{SIR}) to
\begin{equation}
\tau\dot S= -\frac{gS}{g_0}\Big[
(\alpha_X+g_0)(1-S)+(1+\alpha_X)\log(S) \Big]\,,
\label{SIR_reduced}
\end{equation}
which is one dimensional. Integrating equation (\ref{SIR_reduced})
with $g=g(S)$ yields $S=S(t)$, from which $I(t)$ follows
via $\tau\dot I = \big(gS -1\big)I$ and
$R(t)$ from the normalization condition
$S+I+R=1$.

\pdfbookmark[2]{Large control limit of the XI representation}{large_control_limit}
\subsection*{Large control limit of the XI representation}
Expanding equation (\ref{XI}) in $X$, which becomes small
when $\alpha_X\gg1$, one obtains
\begin{equation}
I=\frac{1+\alpha_X}{2g_0}X\left[
2\,\frac{g_0-1}{1+\alpha_X}-X\right] + O(X^3)\,,
\label{XI_expansion}
\end{equation}
which makes clear that the phase-space trajectory becomes
an inverted parabola when infection fractions are small.
As a consequence one finds
\begin{equation}
I \approx \frac{g_0-1}{g_0}\,X + O(X^2)\,,
\label{slope_X}
\end{equation}
which shows that the slope $dI/dX=(g_0-1)/g_0$ at $X\to0$
is independent of $\alpha_X$ and of the normalization procedure
used for $I$ and $X$. The first result was to be expected,
as $\alpha_X$ incorporates the reaction to the outbreak, which
implies that $\alpha_X$ contributes only to higher order. The
dimensionless natural growth factor $g_0$ is hence uniquely
determined, modulo the noise inherent in field data, by measuring
the slope of the daily case numbers with respect
to the cumulative case count.

From equation (\ref{XI_expansion}) one obtains
\begin{equation}
X_{\rm tot}\big|_{\alpha_X\gg1} \approx 2\,\frac{g_0-1}{\alpha_X}\,
\label{X_tot_alpha}
\end{equation}
for the total number of infected $X_{\rm tot}$ in
the large-control limit. In analogy one finds
\begin{equation}
I_{\rm peak}\big|_{\alpha_X\gg1}\approx \frac{(g_0-1)^2}{g_0\alpha_X},
\quad\quad
X_{\rm tot} \approx \frac{2g_0}{g_0-1}\,I_{\rm peak}
\label{I_peak_alpha}
\end{equation}
from equation (\ref{XI}), and in comparison with equation (\ref{X_tot_alpha}).

\pdfbookmark[2]{Time scale asymmetry}{time_scale_asymmetry}
\subsection*{Time scale asymmetry}
From the one-dimensional representation (\ref{SIR_reduced})
of the controlled SIR model one can estimates two
characteristic time scales. For this purpose one considers
an initial relative infection status $f_X X_{\rm tot}$,
with $f_X>0$ and $f_X\ll1$.
\begin{itemize}
\item[--] {\bf run-up}: $T_{\rm up}$, defined as
the time needed to reach the peak when starting from
$X_{\rm start}=f_X X_{\rm tot}$.
\item[--] {\bf run-down}: $T_{\rm down}$, defined as
the time needed to reach $X_{\rm end}=(1-f_X) X_{\rm tot}$,
down from the peak.
\end{itemize}
In general one needs to integrate equation (\ref{SIR_reduced})
numerically. Given that real-world fractional case counts
$X$ are small, $X< X_{\rm tot}\ll1$, one can simplify
(\ref{SIR_reduced}), as for (\ref{XI_expansion}),
obtaining
\begin{equation}
t-t_0 = \frac{\tau}{g_0-1}\log\left(
\frac{X}{(X_{\rm tot}-X)^{2g_0-1}}\right)\,.
\label{time_X_t}
\end{equation}
It follows directly that $T_{\rm down}/T_{\rm up}=2g_0-1$,
as stated in equation (\ref{time_down_up}). For a
pathogen to spread its dimensional growth factor
$g_0$ needs to be larger than unity, compare
Table~\ref{table_CEI}. Going down takes hence
substantially longer than ramping up.

\pdfbookmark[2]{Data availability}{data_availability}
\subsection*{Data availability}
The COVID-19 data examined is publicly accessible
via the  COVID-19 Github repository of the John
Hopkins Center of Systems Science and Engineering\newline
\href{https://github.com/CSSEGISandData/COVID-19}
     {https://github.com/CSSEGISandData/COVID-19}.\newline
Data for the 2015 MERS outbreak in South Korea
is publicly available from the archive of
the World Health organization (WHO),\newline
\href{https://www.who.int/csr/disease/coronavirus_infections/archive-cases/en/
}{https://www.who.int/csr/disease/coronavirus
\_infections/archive-cases/en/}.



\pdfbookmark[1]{Acknowledgments}{Acknowledgments}
\section*{Acknowledgments}
We thank Erik Gros for carefully reading the manuscript,
Andrea Renda and Klaus W{\"a}lde for useful comments and
Angela Capolongo for simulation support. 
We acknowledge financial support from the Horizon 2020 
research and innovation program of the EU under 
grant agreement No. 101016233, H2020-SC1-PHE CORONAVIRUS-2020-2-RTD,
PERISCOPE (Pan European Response to the Impacts of Covid-19 
and future Pandemics and Epidemics)
and from the Fullbright foundation (D.G.).

\pdfbookmark[1]{Author contributions}{contributions}
\section*{Author contributions}
Modeling and theory by C.G. and R.V, data analysis by L.S.,
medical aspects by K.V, economical and political topics by D.G.

\pdfbookmark[1]{Competing interests}{Competing_interests}
\section*{Competing interests}
The authors declare no competing interests.

\pdfbookmark[1]{Additional information}{Additional_information}
\section*{Additional information}
{\bf Correspondence and requests for materials} should be addressed to C.G.

\newpage
\section*{SUPPLEMENTARY INFORMATION}

\renewcommand{\figurename}{{\bf Supplementary Figure}}
\setcounter{figure}{0}

\begin{figure*}[t]
\centerline{
\includegraphics[width=0.9\textwidth]{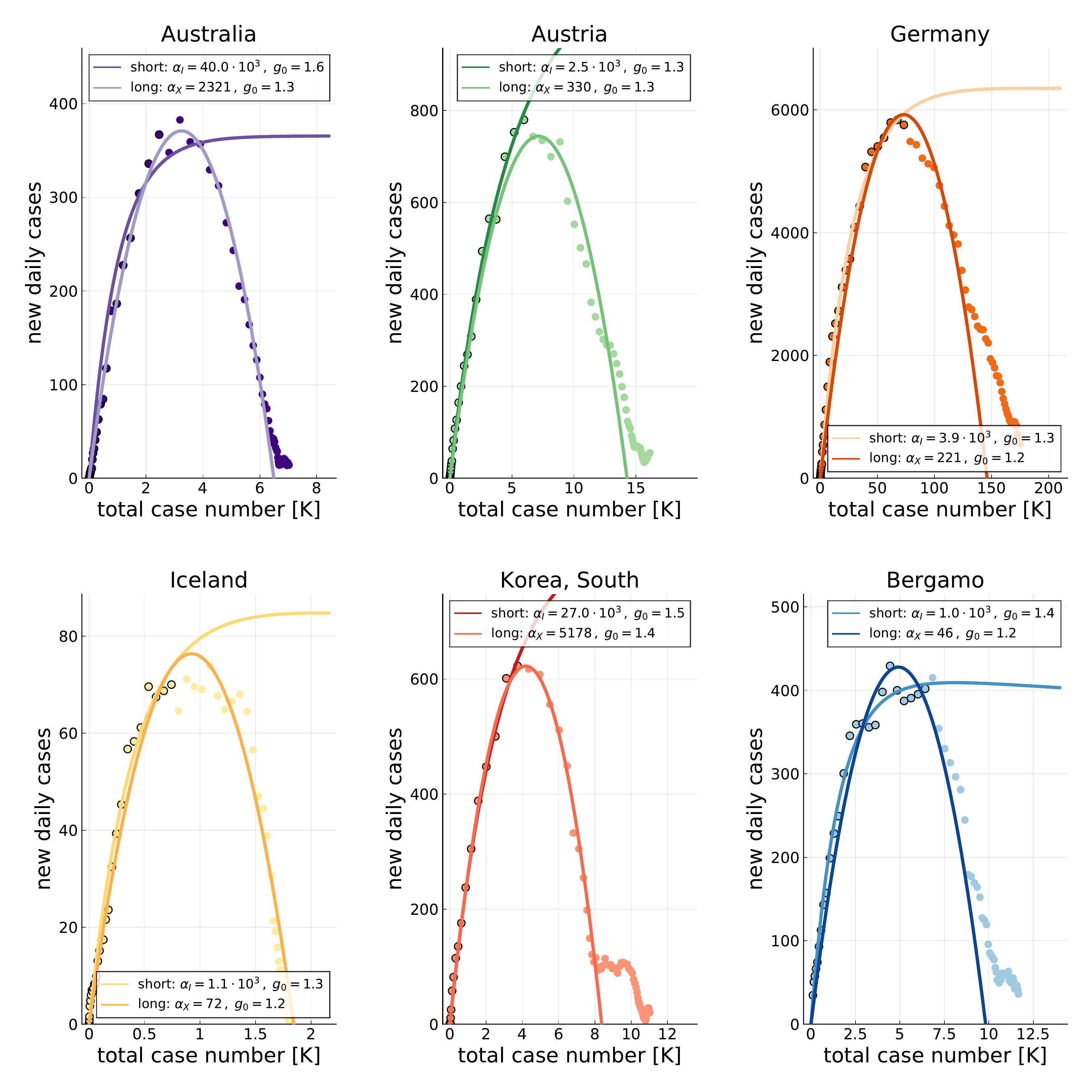}
           }
\caption{{\bf $|$ Short- vs.- long-term containment.}
Fits to COVID-19 data for short-term control,
$(\alpha_I,\alpha_X)=(\alpha_I,0)$ and, for comparison,
for long-term control, $(\alpha_I,\alpha_X)=(0,\alpha_X)$.
The parameter value are given in the respective legends.
The intial slope, which reflect the initial phase of
exponential growth, has been taken as a reference.
}
\label{fig_short_term_fits}
\end{figure*}

\subsection*{Societal reaction to the spread of the disease}

The two strategies investigated here, short-term
and long-term, correspond to reaction patterns
that are observed for the COVID-19
outbreak~\cite{UKpolicyCanged,USpolicyCanged}.
In Supplementary Figure 1 we present some examples.
For both cases, the societal reaction described by
the parameter $\alpha_X$ and $\alpha_I$
can be thought as a sum of two contributions
\begin{equation}
\alpha_X=\alpha_s+\alpha_g,
\end{equation}
where $\alpha_s$ quantifies the
spontaneous reaction by the population and $\alpha_g$ encodes
government interventions. Analogously, such a sum of two contributions
can be made for $\alpha_I$.

The first contribution, $\alpha_s$, takes into account
societal behavioral changes happening when a substantial
fraction of the population spontaneously adopts social
distancing (avoiding hand-shakes, restaurants, cinemas,
etc.), f.i.\ in response to media reports about the severity
of the outbreak. Voluntary social distancing can lead to substantially
reduced restaurants and cinemas attendances even before
governments impose mandatory school closures,
curfews and other drastic measures.\cite{deforche2020behavioral}
An important aspect of the spread of COVID-19 is the distinct
reactions of societies in different countries. In Asia the wearing
of masks becomes a convention, which is likely to correspond to
a higher $\alpha_s$, while such a measure tends to be
resisted by a majority of European populations.\cite{han2020lessons}

The second contribution to the control factor, $\alpha_g$,
captures the role of government interventions. Measures
ranging from forbidding large events, to school closures
and, finally, to lockdowns, become politically possible
when the number of individuals infected increases and
surpasses critical levels. 

The aim of our investigation
is not to evaluate the effectiveness of specific
measures, which has been done elsewhere
\cite{wilder2020can,ferguson2020impact}, but to assess the
dynamical effect of the societal reaction encoded in
the feedback parameter $\alpha_X$,
on the overall evolution of the epidemic.
As most social distancing measures are costly, both
for the economy and overall well-being\cite{every2020psychological}, 
it is reasonable to
assume that their strictness is increased only when
necessary, viz in relation to the severity of the
outbreak. The latter can be measured  either
by the number of current cases, $I_t$, or by the cumulative
case count $X_t$.

The inverted U shape of the total cost of the virus as a function of $\alpha_X$
has one important corollary: the 'laissez faire' equilibrium is not the optimum
for society. If the government abstains from action leaving it to societal
reaction to dampen the peak,  the spread of the disease would be limited only by
$\alpha_s$, which might bring society close to the hump of the total cost curve.
Strong government action, i.e.\ a high value of $\alpha_g$ could then push the
path to the other side of the hump resulting in lower costs.
In other words,
relying only on individual reaction which aims at lowering the risk to oneself,
would be sub-optimal. This is of course a general result for all contagious
diseases \cite{althouse2010public,roberts2010guide} but we confirm it
accounting explicitly for the cost of the measures needed to protect public
health.

\subsection*{Detailed costs of controlling the COVID-19 pandemic}

In what follows we present a detailed estimation
of the costs of controlling the COVID-19 pandemic
given in GDP per capita (GDP$_{\rm p.c.}$) to ensure comparability
across countries. 

Four elements dominate the cost structure:  (i) The
working time lost due to an infection, (ii) the direct
medical costs of infections, (iii) the value of life costs, and
(iv) the cost related to `social distancing'. The first three
are medical or health-related.

\subsection*{Health costs, loss of working time}

A first direct impact of a wave of infections is that a fraction
of the population cannot work. Based on the Diamond Princess
data~\cite{russell2020estimating}, where the entire population
was tested, we estimate that only half of the infected develop
symptoms that require them to stay home for a one- to two-week
period and an additional two-week period
until they are no longer contagious. About 20\% of the infected
(or 40\%  of those with symptoms) develop stronger
symptoms requiring one additional period of absence from work
\cite{raoult2020coronavirus}. To be conservative, we assume that
there are no severe cases or deaths among the working age.
This results in a reduction in the work force per year (52 weeks)
of around
$(0.3\times2 + 0.2\times3)\times(2/52)= 2.4/52 = 0.05$, 
for every 1\% of the population infected.

\subsection*{Medical costs, treatment, hospitalization}

Hospitalization rates and costs of hospital
treatment for COVID-19 vary enormously across across countries.
But it is estimated that about 20\%  \cite{covid2020severe} 
of the infected individuals
require some sort of hospitalization. 
A recent large scale survey of the literature\cite{tzotzos2020incidence}
shows that about one fourth of them (around 5\% of the infected)
need intensive care and roughly on sixth dying.
We adopt a more conservative fatality rate among
the hospitalized of 1\% \cite{russell2020estimating}.

As a comparison, we note that an average influenza season leads 
to a hospitalization of about 0.12\% of the US population
\cite{CDC_flue}; and one fourth of them require intensive
care, with one twentieth (0.13\% of all infected) dying
\cite{russell2020estimating}. Averaged over the 2010-17,
of the order of 35 thousand influenza-related deaths per
year have been registered in the US, one tenth of the over
300 thousand COVID-19 related fatalities registered in 2020. 

Costs: Intensive care with ventilation is the most costly form of
life saving in hospital care\cite{tzotzos2020incidence}.  
In the US, the cost of two
weeks of an intensive care unit is equivalent to about 
1 year (100\%) of GDP$_{\rm p.c.}$~\cite{chen2020medical}. 
For Covid-19 patients in the US, it has been reported that average
hospitalisation costs per case amount to over 70 
thousand USD, or 115\% of (annual) GDP per capita~\cite{fairhealth2020}.

In Germany, which might be typical of the rest of Europe, 
the cost of two weeks of intensive care appears to be
lower, around 20,000 Euro, or roughly 60\% of 
GDP$_{\rm p.c.}$~\cite{Martin2008}. We use the German 
parameters for a conservative estimate of medical costs.
The cost of general hospitalization for two weeks is assumed 
to be 12,000 Euro, and equivalent to about 30\%
of GDP$_{\rm p.c.}$. With two weeks of general hospitalization 
and two weeks of intensive care for severe cases, this results 
in a medical cost of $(0.05\times0.6 + 0.05\times0.3 
+ 0.15\times0.3 = 0.09)$, that is of 9\% GDP$_{\rm p.c.}$.

\subsection*{Value of lives lost}

Third, the cost of premature death through the disease
represents the most difficult element to evaluate in financial terms.
We will show below that our central results
remain valid even without assigning a monetary value to lives lost,
but since major contributions~\cite{flatteningCurve2020}
are based on an evaluation of the economic value of lives lost,
we show how this point can be incorporated into our framework.
There are two ways to attribute a monetary value on a life saved or lost.
The first, mentioned in the main text, is based on the concept of a
Value of Statistical Life (VSL), which is commonly used in
the impact assessment of public policy that aims at
lowering the probability of a premature
death~\cite{renda2013assessing}. A typical application scenario
for VSL is the case when the probability of death is very low
(e.g.\ car accidents), but could be lowered even more (seat belts).
For COVID-19, a high-death epidemic, we prefer a medical-based
approach, which allows us to produce conservative estimates.
VSL arrives in contrast often at much higher values, up to millions
of Euro or Dollars~\cite{USDA}.
Putting a monetary value on lives saved is unavoidable in medical
practice that is confronted with the problem of selecting
the procedures to be used to prolong life - a situation that
arises for many patients infected by the Coronavirus
under intensive care. 

The literature dealing with the cost
of medical procedures finds a central range of between
100,000 and 300,000 Dollars per year of
life lost~\cite{Neumann2014,HST}.
Given the current US GDP$_{\rm p.c.}$, these values translate
into a range of 1.5 to 4 years of GDP$_{\rm p.c.}$. Cutler and
Richardson~\cite{NBERw6895} argue for a value equivalent to
three times GDP$_{\rm p.c.}$. We use the lower bound of
this range (i.e. 1.5 times GDP$_{\rm p.c.}$) 
for most of our simulations, 
which might thus under-estimated 
the value of lives saved through social distancing restrictions.

What remains to be determined is the number of years lost 
when a Corona patient dies.

On the cruise
ship Diamond Princess~\cite{russell2020estimating,rocklov2020covid} which
served almost as a laboratory, the average age at death was 76 years. Cruise
passengers tend to have fewer acute health conditions than the general
population, thus rendering the co-morbidity argument less prominent. The
remaining life expectancy (weighted by the difference incidence by sex)
would thus be 11 years.
This implies that the economic value of the premature deaths
should be equal to about 11 times the loss for one year of life saved
(potentially higher for most European countries which tend to have a higher
life expectancy). 

For each 1\% of the population the value of lives lost
would thus be equal to $0.01\times11\times$ the nominal value of one
year of life.

The value of life can be measured in terms
of multiples of GDP$_{\rm p.c.}$, which allows to
write the sum of the three types of health or medical costs (loss
of working time, hospitalization and value of lives lost)
as a linear function of the percentage of the population infected:
$$
c_t^{\rm med} = kI_t
$$
with a proportionality factor $k$ being equal to the sum of the
three contributions.  Scaling $k$ with the
GDP$_{\rm p.c.}$ allows for an application and comparison
across countries. Using the lower bound of the central range
yields then the following calibration of the medical costs:
\begin{eqnarray}
\nonumber
(0.05 + 0.09 + 0.01\times1.5\times11)\times \mbox{GDP}_{\rm p.c.}
&=& \\
0.305 \times \mbox{GDP}_{\rm p.c.} &&
\label{costs_medical}
\end{eqnarray}
The upper bound for the value of $k$ would be substantially higher:
$(0.05 + 0.09 + 0.01\times4\times11)\times \mbox{GDP}_{\rm p.c.}=
0.58 \times \mbox{GDP}_{\rm p.c.}$. For the numerical calculations
we will use the conservative estimate  $k=0.305$ in terms of GDP$_{\rm p.c.}$.

If we only consider the direct medical costs consisting of loss
of working time and hospitalization, without including the value
of lives lost, the proportionality factor in equation~(\ref{costs_medical})
reduces to $k=0.14$ in terms of GDP$_{\rm p.c.}$.

\subsection*{Medical costs over the lifetime of the epidemic}

The cost estimates discussed so far, $c_t^{\rm med}$,
refer to the per-period cost of the currently infected.
For the total cost over the entire endemic we need to
calculate the discounted sum of all $c_t^{\rm med}$ over time.
Given that a period corresponds to about two weeks, we neglect
discounting, which would make little difference
even if one uses a social discount rate of 5\% instead
of using market rates (which may be negative).
The total medical costs over the course of the endemic
can be written as the simple sum of the cost per unit of time:
\begin{equation}
C^{\rm medical} = \sum_{I_{t}>I_{\rm min}} c_t^{\rm med}=kX_{tot}\,.
\label{costs_medical and value of life}
\end{equation}
The epidemic is considered to have stopped when the fraction
of new infections $I_t$ falls below a minimal value,
$I_{\rm min}$.

Using the conservative estimate (low value of life)
$k$ = 0.305 it is straightforward to evaluate the total
cost of a policy of not reacting at all to the spread
of the disease, which would lead in the end to
$X_{\rm tot}=0.94$. A hands-off policy would therefore
lead to medical costs of over 28\% of GDP.

In absolute terms the cost of a
policy of doing nothing would amount to 1000 billion
Euro for a country like Germany. For the US  the sum
would be closer to 5 Trillion of Dollars (25\% of a
GDP of 20 Trillion of Dollars). As it would not be
possible to ramp up hospital capacity in the short time
given the rapid spread of the disease, the cost would be
in reality substantially higher, together with the death 
toll~\cite{flatteningCurve2020,ferguson2020impact}.
We abstract from the question of medical capacity (limited
number of hospital beds) because we assume that society
would react anyway as the virus spreads, thus limiting the peak,
and, second, we are interested in the longer term implications
of different strategies and not just in their impact on the
short-term peak.

We note that even concentrating only on the direct
medical cost and working time lost ($k$=0.14)
a policy of letting
the epidemic run its course through the entire population
would lead to losses of working time and hospital treatment
of over 13\% of GDP (94\% of 14\%). By comparison, total
health expenditure in most European countries amounts in
normal times to about 11\% of GDP~\cite{statista}. Even
apart from ethical considerations, to avoid or not
potentially hundreds of thousands of premature deaths,
there exists thus an economic incentive to slow
the spread of the COVID-19 virus.

Given the somewhat contentious nature of the value
of lives lost, we present in the middle panel
of Fig. 4 of the main text the
medical cost estimates (as a proportion of GDP) without
including the value of life costs (results with including
the value of life costs are shown in the main text).
As shown in the figure, increasing $\alpha_X$ leads to a
lower medical cost because the percentage of the population
infected will be lower. The difference between short-term
and long-term control increases for higher values
of $\alpha_X$. At these $\alpha_X$ values the medical cost
over the entire endemic would be lower because the overall
fraction of infected population is lower. For a strongly
reactive society and policy i.e.\ for $\alpha_X\gg1$ (and
the case of long-term control), an explicit solution
for the total health cost is given by,

\begin{equation}
C^{\rm medical} = kX_{\rm tot}\big|_{\alpha_X\gg1} \approx 2k\,\frac{g_0-1}{\alpha_X}\,
\end{equation}
which implies that the total health or medical costs
are inversely proportional
to the strength of the policy reaction parameter. Draconian
measures from the start, i.e.\ with $\alpha_X$ going towards 
infinity, reduce the medical costs to close to zero - 
irrespective of whether one adds the value of lives lost. 
This can be seen in Fig.\,3 and Fig.\,4 of the main text,
where the medical cost (over the entire epidemic) starts 
for $\alpha_X=0$ at values close to $k$ because without 
any societal reaction 94\% of the population would get 
infected and with increasing $\alpha_X$ the medical costs 
decline monotonously.

\subsection*{Social distancing costs}

The economic costs of imposing social distancing on a wider population are at
the core of policy discussions and drive financial markets.  As mentioned
above, social distancing can take many forms; ranging from abstaining from
travel or restaurant meals to government interventions
enforcing lockdowns, quarantine, closure of schools, etc.  This cost is more
difficult to estimate.  However, a rough estimate is possible if one takes into
account that most economic activity involves some social interactions. Limiting
social interaction thus necessarily reduces economic activity.
This suggests that the economic cost of the social distancing
described in equation~(2) of the main text should increase with
the reduction in the transmission rate described by
$g$.

Without any social distancing, $\alpha_X=0$, 
the economy would not be affected by
the spread of the virus. Stopping all economic social interactions would bring
the economy to a halt, but the reproduction rate of the virus would also go
close to zero (Eichenbaum {\it et al.}~\cite{Eichenbaum2020}
make a similar assumption). We thus posit that the (per-time unit)
social-distancing economic cost ${c}_t^{\rm s}$  is proportional to
the reduction in the transmission rate. The total economic costs
$C^{\rm social}$
can be written as the sum of ${c}_t^{\rm s}$:
\begin{equation}
C^{\rm social} = \sum_{I_{t}>I_{\rm min}} \mbox{c}_t^{\rm s},
\quad\quad
\mbox{c}_{t}^{\rm s} =
m\left[1-\frac{\rho_t}{\rho_0}\right]\frac{2}{52}
\label{supp_costs_social}
\end{equation}
considering here the notation of the discrete-time controlled
SIR model (equation (8)).
The key question is the factor of proportionality, $m$, which
links the severity of social distancing to the reduction in
economic activity. Popular attention has focused on services
linked directly to social contact. There exist indeed selected
sectors which will completely shut down under a lockdown.
However, these sectors (tourism, non-food retail, etc.) account
for a limited share of the economy (less than 10\% for most
countries). Expenditure for food is actually little affected
since even under the most severe lockdown, grocery shopping is
still allowed and families must consume more food at home as
they cannot go out to restaurants.

The manufacturing sector is less affected by social distancing
than the service sector because in modern factories workers are
scattered over a large factory floor, making it relatively easy
to maintain production while maintaining the appropriate distance
between workers. Moreover, some sectors, e.g.\ finance, can work
online with only a limited effect on productivity. The widespread
impression that the entire economy stops under a lockdown is thus
not correct. The drastic measures adopted in China illustrate this
proposition: when all non-essential social interactions were forbidden,
industrial production and retail sales fell by 'only' 20-25\%~\cite{IMF}
while the reproduction factor went from 3 to 0.3, a fall by a factor
of ten. Using this experience we calibrate the parameter $m$ at 0.25.
The projections of the International Monetary Fund, a loss of output
of about 8\% for severe lockdowns lasting one quarter
\cite{IMFoutlook2020}, \cite{alvarez2020simple} confirm this order of magnitude.

A reduction in the reproduction factor $\rho_t$ to one
tenth its normal epidemiological value of $\rho_0$ would
thus lead to a loss of GDP of 25\% for the time period
during which the restriction or social distancing measures
are in place. This would imply that an abrupt shutdown of
the economy to 25\% of its capacity for 12 weeks, or 6
incubation periods would cost about 0.25$\times$(12/52),
or about 6\% of annual GDP. A reduction of GDP by 6\%
represents a recession even deeper than the one
which followed the financial crisis of 2009.
This is compatible with current forecasts of zero GDP
growth in China in 2020
(relative to a baseline of 5-6\% before the crisis).
But even such a large cost in terms of output foregone would be below the
medical cost arising from herd immunity\cite{rowthorn2020cost}.  
Even apart from ethical
considerations, it would thus appear to make sense to accept a temporary shut
down of parts of the economy to avoid the huge medical costs.

A first result is thus that if one compares two extremes:
letting contagion run its course (herd immunity)
or draconian measures, the social costs are lower in 
the second case. Small changes to the key parameters, $k$
and $m$, might change the exact values of the costs in
terms of overall magnitude, but the ranking appears robust.

We do not consider separately the fiscal cost, i.e.\ the cost
for the government to save millions of enterprises from
bankruptcy and ensure that workers have a replacement income
when they get laid off. This cost to governments is a transfer
within the country from one part of society (tax payers) to those
who suffer most under the economic crisis.

A key issue in the discussion on the economic cost of social distancing
is the question about how long these measures need to be maintained. 
It is sometimes
argued that the cost of a policy of social distancing would be unacceptably
high because the measures could not be relaxed until the virus had been totally
eradicated.  However, this pessimism is not
warranted by the success of a strategy of `testing and tracing' implemented in
some countries (mainly those which had experienced SARS).  Such a strategy is,
of course, only possible if the starting number of infections is low enough to
allow for individual tracing.

We thus make the assumption that when the number of active cases falls below a
certain threshold, the costly measures of general social distance
containment are no longer needed and can be substituted by pro-active
repeated testing coupled to quick follow-up of the remaining few cases which
are quarantined and whose contacts are quickly traced.  In this case the
resulting economic cost is  assumed to fall away.  The experience of 
Taiwan\cite{wang2020response}, Korea\cite{park2020contact}
and Japan suggests that when the infected are less than one per 100,000,
general social distancing is no longer required (assuming mass testing has been
adopted in the meantime so that the infections can be accurately measured).

\bigskip\noindent{\bf Parameter updating}\newline
The estimates on which our results are based will have to
be updated when actualized COVID-19 data is available
in the future. The WHO-China Joint Mission Report
suggests a $\rho_0$ ($g_0$ in the continuous-time representation)
per infected of $2-2.5$~\cite{mission2020report}
(in units of the disease duration),
while we use the figures from Liu
{\it et al.}~\cite{liu2020reproductive}, who predict
a reproduction factor of around three. The numbers for the forecast
of health costs are derived in part from the Diamond Princess data
\cite{russell2020estimating}, for which the population was comparatively
healthy. The statistics for symptoms requiring the absence from work
may therefore in reality be somewhat higher. 

The hospitalization
and mortality rate are estimated with a substantial uncertainty,
due to the high numbers of unregistered and untested infections.
Early studies based on official data from
China~\cite{nishiura2020rate,read2020novel}
estimated that the number of actual infections may be
between 10 to 20 times higher than the
number of detected infections.
However, serological test in e.g. Austria
suggest only a factor of 3~\cite{Austria}.
The continuing screening of blood samples
via the US, 
Nationwide Commercial Laboratory Seroprevalence Survey,
showed in September 2020 an average undercounting
factor of 2.6, as opposed to
our assumption of 2.
Leaving possibly lower, but still substantial
true hospitalization and mortality rates for COVID-19.

A strong age gradient has been observed for the case fatality rate 
of COVID-19 by age\cite{signorelli2020age},
which could be logistic\cite{gao2020logistic},  
and there are large variations across countries\cite{hoffmann2020older}.
Moreover, one has to take into account that while case fatality
rates are much lower for the younger, they are represent a larger 
share of the population and their life expectancy 
is also higher (e.g. over 20 years for the 60 years old).  
These two factors tend to give more weight to the younger age brackets,
leading to resulting parameter estimates similar to ours
\cite{gros2020great}. 

One of our main goals has
been the introduction of a generic framework, which can be updated by
further advances in the accuracy of estimates while still presenting
specific results with the data available at this time.

\bigskip\noindent{\bf Relation to further studies}\newline
A range of determining factors have been examined
for the ongoing COVID-19 epidemic, in particular
the effect of quarantine \cite{peng2020epidemic} and
that community-level social distancing may be more
important than the social distancing of individuals
\cite{siegenfeld2020eliminating}. An agent-based model
for Australia found, in this regard, that school closures
may not be decisive \cite{chang2020modelling}.
Microsimulation models suggest, on the other hand, that
a substantial range of non-pharmaceutical interventions
might be needed for an effective containment of the COVID-19
outbreak \cite{ferguson2020impact}.

We also note attempts to derive disease
transmission rates from economic principles of behavior
\cite{Eichenbaum2020}, which would allow to measure the
cost of the Corona pandemic under different policy settings.
Another strand of the literature takes the pandemic as given,
and as the basis for scenarios for the economic impact and
for the financial-market volatility
\cite{atkeson2020will,baldwin2020economics}.


\end{document}